\documentclass[twocolumn,usenatbib]{mn2e}
\usepackage{amsmath}
\usepackage{epsfig}
\usepackage{capt-of}

\begin{document}

\title{Chaotic and regular motion around generalized Kalnajs discs}

\author[J. Ramos-Caro, F. L\'opez-Suspes and G. A. Gonz\'alez]
{Javier Ramos-Caro\thanks{E-mail: javiramos1976@gmail.com},
Framsol L\'opez-Suspes\thanks{E-mail: framsol@gmail.com} and
Guillermo A. Gonz\'alez\thanks{E-mail: gonzalez@gag-girg-uis.net} \\
Escuela de F\'isica, Grupo de Investigaci\'on en Relatividad y Gravitaci\'on,
Universidad Industrial de Santander,\\
A. A. 678, Bucaramanga, Santander, Colombia}

\maketitle

\begin{abstract}

The motion of test particles in the gravitational fields generated by the first
four members of the infinite family of generalized Kalnajs discs (\cite{GR}), is
studied. In first instance, we analyze the stability of circular orbits under
radial and vertical perturbations and describe the behavior of general
equatorial orbits and so we find that radial stability and vertical instability
dominate such disc models. Then we study bounded axially symmetric orbits by
using the Poincar\'e surfaces of section and Lyapunov characteristic numbers and
find chaos in the case of disc-crossing orbits and completely regular motion in
other cases.

\end{abstract}


\section{Introduction}

A fact usually assumed in astrophysics is that the main part of the mass of a
typical spiral galaxy is concentrated in a thin disk (\cite{BT}). Accordingly,
the study of the gravitational potential generated by an idealized thin disk is
a problem of great astrophysical relevance and so, through the years, different
approaches has been used to obtain such kind of thin disk models(see Kuzmin
(\citeyear{KUZ}) and Toomre (\citeyear{T1,T2}), as examples). A simple method to
obtain the surface density, the gravitational potential and the rotation curve
of thin disks of finite radius was developed by \cite{HUN1}, the simplest
example of disk obtained by this method being the Kalnajs disk (\cite{KAL}),
which can also be obtained by flattening a uniformly rotating spheroid
(\cite{WM,BR,BB}). In a previous paper (\cite{GR}) we use the Hunter method in
order to obtain an infinite family of thin disks of finite radius, an infinite
family of generalized Kalnajs discs with a well behaved surface mass density.

Closely related with the above study is the analysis of the motion of test
particles in the gravitational field generated by such disklike distributions of
matter. In particular, the study of orbits in the equatorial plane is of clear
astrophysical relevance due to its relation with the dynamics of intergalactic
stellar motion or the flow of particles in accretion disks around black holes.
Also, a knowledge of the disc's internal kinematics is relevant for its
subsequent statistical analysis, and the study of external particles motion help
us to understand the behavior of stars belonging to the galaxy's remaining
component (for example, the halo). Many of them cross back and forth through the
disc, experiencing a fairly abrupt change in the gravitational force field. This
fact gives rise to a large variety of chaotic and regular orbits, as it was
pointed out by Hunter (\citeyear{Hunter}), in the case of Kuzmin-like discs, and
Martinet et al (\citeyear{Martinet1}; \citeyear{Martinet2};
\citeyear{Martinet3}; \citeyear{Martinet4}), in the case of Schmidt's models
(Schmidt \citeyear{Schmidt}).

In agreement with the above considerations, in this paper we shall focus on the
kinematics around the generalized Kalnajs discs, introduced by Gonz\'alez \&
Reina (\citeyear{GR}). They form an infinite family of axially symmetric finite
thin discs, whose first member is precisely the well-known Kalnajs disc
(\citeyear{KAL}).  The paper is organized as follows. First, in section
\ref{sec:kal}, we present a summary of the main aspects of the generalized
Kalnajs discs, the surface densities, the gravitational potential and the motion
equations. Then, in section \ref{sec:equ} we shall focus on equatorial orbits,
i.e. $z=0$ trajectories ($\xi=0$, $\eta=\sqrt{1-R^{2}/a^{2}}$ inside the disc
and $\xi=\sqrt{R^{2}/a^{2}-1}$, $\eta=0$ outside). At first we shall study the
stability under radial and vertical perturbations  of circular orbits, then
examining the principal features and conditions of general equatorial orbits. In
the next section, section \ref{sec:cros}, we present numerical solutions of
(\ref{em}) describing some representative Poincar\'e surfaces of section. As it
is expected, we find chaotic sections for disk-crossing orbits and regular
sections for other cases. Some meridional plane orbits are plotted and the
Lyapunov characteristic numbers (LCN) are calculated. Finally, in section
\ref{sec.con}, we summarize our main results.

\section{The generalized Kalnajs discs}\label{sec:kal}

In this paper we shall focus on the kinematics around generalized Kalnajs discs,
introduced by Gonz\'alez \& Reina (\citeyear{GR}). They form an infinite family
of axially symmetric finite thin discs, whose first member is precisely the
well-known Kalnajs disc (\citeyear{KAL}). The mass surface density of each model
(labeled with the positive integer $m$) is given by
\begin{equation}
\Sigma_{m}(R) = \frac{(2m+1)M}{2\pi a^{2}} \left[ 1 - \frac{R^{2}}{a^{2}}
\right]^{m-1/2}, \label{densidad}
\end{equation}
where $M$ is the total mass and $a$ the disc radius. Such mass distribution
generates an axially symmetric gravitational potential, that can be written in
terms of Legendre polynomials $P_{n}$ and second kind Legendre functions $Q_{n}$
as
\begin{equation}
\Phi_{m}= -\sum_{n = 0}^{m}C_{2n}P_{2n}(\eta)i^{2n + 1}Q_{2n}(i\xi).\label{dk}
\end{equation}
Here, $-1 \leq \eta \leq 1$ and $0 \leq \xi < \infty$ are spheroidal oblate
coordinates, related to the usual cylindrical coordinates $(R,z)$ through the
relations
\begin{equation}
R^{2} = a^{2}(1 + \xi^{2})(1 - \eta^{2}),\qquad z =a\eta\xi.\label{tco}
\end{equation}
The constants $C_{2n}$ appearing in (\ref{dk}) are given by
$$
C_{2n}= \frac{M G}{2 a} \left[\frac{\pi^{1/2} (4n+1) (2m+1)!}{2^{2m} (2n+1) (m - n)! \Gamma(m +
n + \frac{3}{2} ) q_{2n+1}(0)} \right],
$$
where $q_{2n} (\xi) = i^{2n+1} Q_{2n} (i\xi)$ and $G$ is the gravitational
constant. Now, due to the presence of the term $(m - n)!$ at the denominator,
all the $C_{2n}$ constants vanish for $n > m$.

According to (\ref{dk}), the gravitational potentials corresponding to the first
four members are given by
\begin{subequations}\begin{align}
\Phi_{1}(\xi,\eta) &= - \frac{MG}{a} [ \cot^{-1}\xi  + A (3\eta^{2} - 1)],
\label{eq:4.22}   \\
\Phi_{2}(\xi,\eta) &= - \frac{MG}{a} [ \cot^{-1} \xi + \frac{10 A}{7}
(3\eta^{2} - 1) \nonumber \\
 &  \quad \quad  + \ B ( 35 \eta^{4} - 30 \eta^{2} + 3)], \label{eq:4.23}  \\
\Phi_{3}(\xi,\eta) &= - \frac{MG}{a} [ \cot^{-1} \xi + \frac{5 A}{3} (3
\eta^{2} - 1)  \nonumber \\
    & \quad \quad + \ \frac{9 B}{11} (35 \eta^{4} - 30 \eta^{2} + 3)
\nonumber   \\
&  \quad \quad  + \ C (231 \eta^{6} - 315 \eta^{4} + 105 \eta^{2} - 5) ], \\
\Phi_{4}(\xi,\eta) &= - \frac{MG}{a} [ \cot^{-1} \xi + \frac{20 A}{11} (3
\eta^{2} - 1)  \nonumber \\
    & \quad \quad + \ \frac{162 B}{143} (35 \eta^{4} - 30 \eta^{2} + 3)
\nonumber   \\
&  \quad \quad  + \ \frac{4C}{11} (231 \eta^{6} - 315 \eta^{4} + 105 \eta^{2} -
5) \nonumber \\
&\quad \quad +D(6435 \eta ^8-12012 \eta ^6+6930 \eta ^4\nonumber \\
&\quad \quad-1260 \eta ^2+35)],
\end{align}\end{subequations}
with
\begin{subequations}\begin{align}
A &= \frac{1}{4} [(3\xi^{2} + 1) \cot^{-1} \xi - 3 \xi ], \\
B &= \frac{3}{448} [ (35 \xi^{4} + 30 \xi^{2} + 3) \cot^{-1} \xi - 35 \xi^{3} -
\frac{55}{3} \xi ], \\
C &= \frac{5}{8448} [ (231 \xi^{6} + 315 \xi^{4} + 105 \xi^{2} + 5) \cot^{-1}
\xi  \nonumber \\
& \quad \quad - 231 \xi^{5} - 238 \xi^{3} - \frac{231}{5} \xi ], \\
D &= \frac{7}{2342912} [\left(6435 \xi ^8+12012 \xi ^6+6930 \xi
   ^4\right.\nonumber \\
   &\quad \quad\left.+1260 \xi ^2+35\right) \cot ^{-1}\xi-6435 \xi ^7-9867 \xi
   ^5\nonumber \\
   &\quad \quad-4213 \xi^3-\frac{15159 \xi }{35}].
\end{align}\end{subequations}
We restrict our attention to these four members. The kinematics corresponding to
the remaining models ($m\geq 5$) is easily inferred from the features
characterizing $m=1,2,3,4$. In all the calculations concerning with such models
we choose $a=M=G=1$, without loss of generality.

Since each ${\Phi}_{m}$ is static and axially symmetric, the specific energy $E$
and the specific axial angular momentum $\ell$ are conserved along the particle
motion. This fact restricts such motion to a three dimensional subspace of the
$(R,z,V_{R},V_{z})$ phase space. By defining an effective potential
$\tilde{\Phi}_{m}$ as
\begin{equation}
\tilde{\Phi}_{m} = \Phi_{m} + \frac{\ell^{2}}{2R^{2}},
\end{equation}
the motion will be determined by the equations (Binney \& Tremaine
\citeyear{BT})
\begin{eqnarray}
\dot{R}&=&{V}_{R},\qquad \dot{z}={V}_{z}, \nonumber\\&&\nonumber\\
\qquad\dot{V}_{R}&=&-\frac{\partial \tilde{\Phi}_{m}}{\partial
R},\qquad \dot{V}_{z}=-\frac{\partial \tilde{\Phi}_{m}}{\partial
z},\label{em}
\end{eqnarray}
together with
\begin{equation}
E = \frac{1}{2}({V}_{R}^{2}+{V}_{z}^{2}) + \tilde{\Phi}_{m}.
\end{equation}

\section{Equatorial orbits}\label{sec:equ}

The equilibrium points of the autonomous system (\ref{em}) are
$V_{R}=V_{z}=z=0$, $R=R_{c}$, where $R_{c}$ must satisfy the equation
\begin{equation}
\left(\frac{\partial \tilde{\Phi}_{m}}{\partial R}\right)_{(R_{c},0)}=
-\frac{\ell^{2}}{R_{c}^{3}}+\left(\frac{\partial \Phi_{m}}{\partial
R}\right)_{(R_{c},0)}=0,\label{circular}
\end{equation}
that is the condition for a circular orbit in the plane $z=0$. In other words,
the equilibrium points of (\ref{em}) occur when the test particle describes
equatorial circular orbits of radius $R_{c}$, specific axial angular momentum
given by
\begin{equation}
\ell_{c}=\pm \sqrt{R_{c}^{3}\left(\frac{\partial \Phi_{m}}{\partial
R}\right)_{(R_{c},0)}}\:\:,\label{Lz}
\end{equation}
and specific energy
\begin{equation}
E=\tilde{\Phi}_{m}(R_{c},0),\label{Ecircular}
\end{equation}
where the subscript $c$ in $\ell_{c}$ indicates that we are dealing with
circular orbits.

\begin{figure}
\epsfig{width=2.75in,file=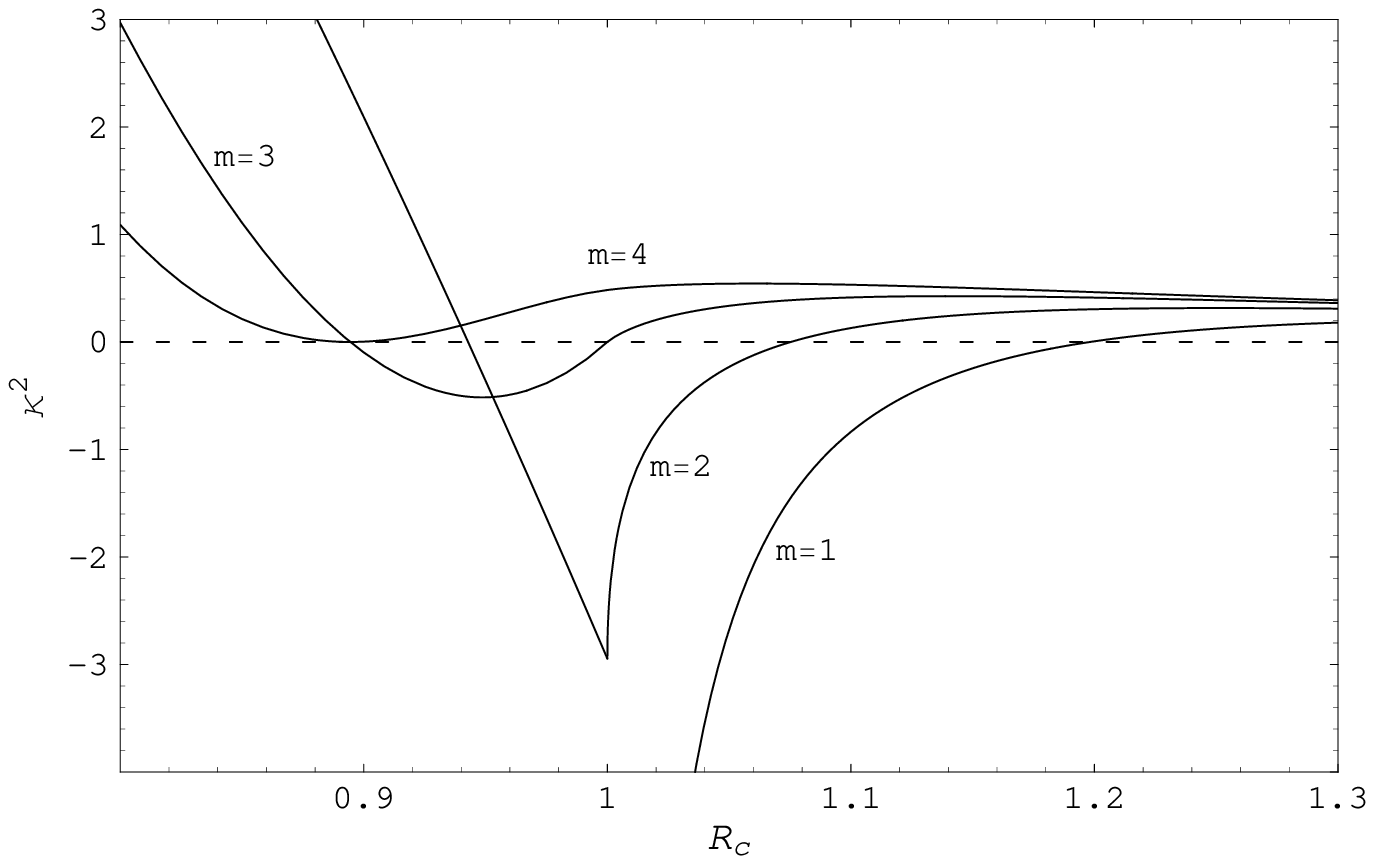}\\
  \caption{Behavior of $\kappa^{2}$ for $m=1,2,3,4$. Values of $R_{c}$ such
  that this function is below the dashed line, corresponds to circular orbits
  that are unstable under radial perturbations.}\label{epiciclica}
\epsfig{width=2.75in,file=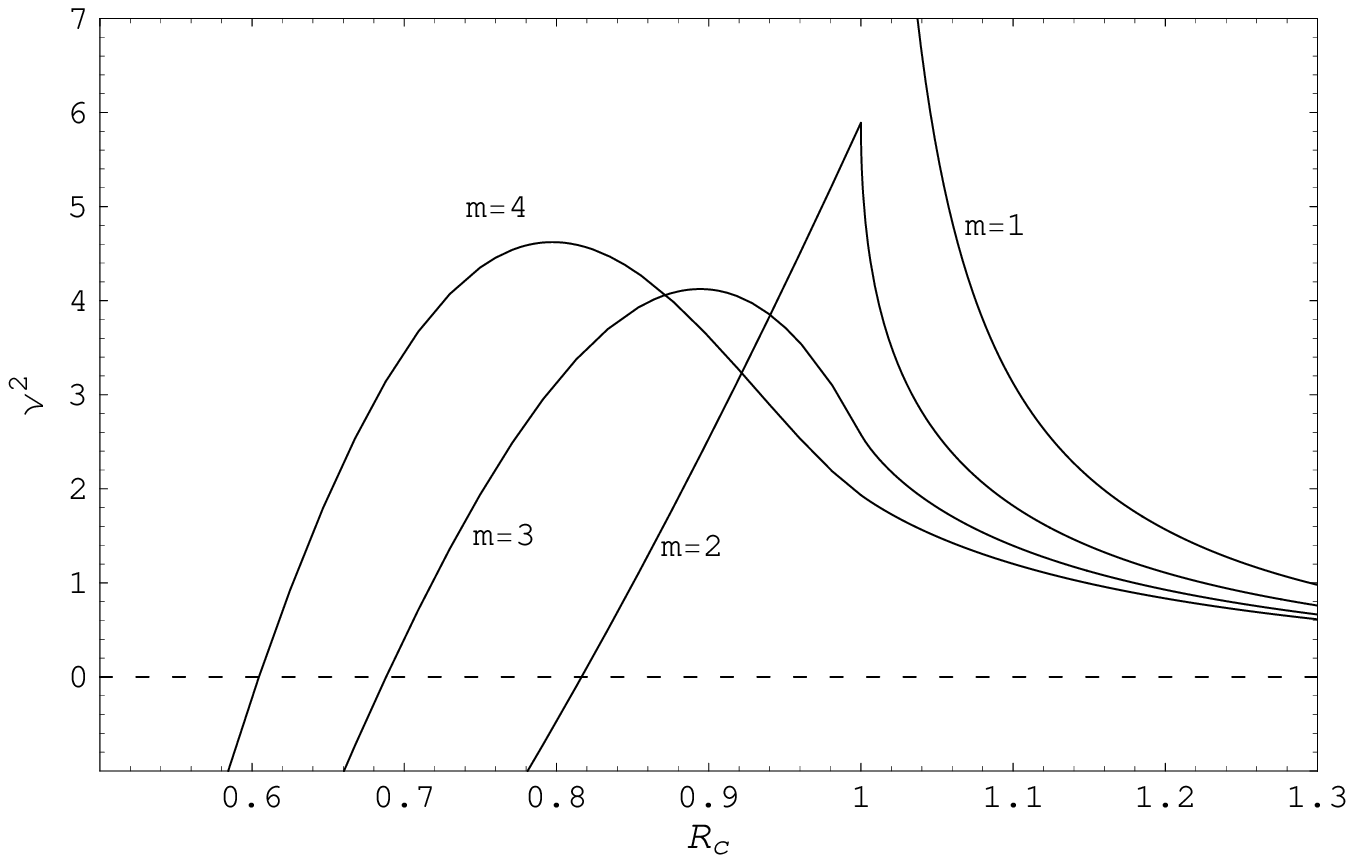}\\
  \caption{Behavior of $\nu^{2}$ for $m=1,2,3,4$. Values of $R_{c}$ such that
  this function is below the dashed line, corresponds to circular orbits that
  are unstable under vertical perturbations.}\label{vertical}
\end{figure}

In order to study the stability of these orbits under small radial and vertical
($z$-direction) perturbations, we analyze the nature of quasi-circular orbits.
They are characterized by an epicycle frequency $\kappa$ and a vertical
frequency $\nu$, given by (Binney \& Tremaine \citeyear{BT})
\begin{equation}
\kappa^{2}=\left(\frac{\partial^{2} \tilde{\Phi}_{m}}{\partial
R^{2}}\right)_{(R_{c},0)}, \:\:\:\:\:\:\:\:\:\:
\nu^{2}=\left(\frac{\partial^{2} \tilde{\Phi}_{m}}{\partial
z^{2}}\right)_{(R_{c},0)}.\label{epicicle}
\end{equation}
This means that by introducing (\ref{Lz}) in the second derivatives of
$\tilde{\Phi}_{m}$ we obtain $\kappa^{2}$ and $\nu^{2}$ as functions of $R_{c}$.
Values of $R_{c}$ such that $\kappa^{2}>0$ and (or) $\nu^{2}>0$ corresponds to
stable circular orbits under small radial and (or) vertical perturbations,
respectively. Otherwise we find unstable circular orbits. The case $m=1$
presents radial stability in the range $0\leq R_{c}\leq 1$ ($\kappa^{2}=3\pi$)
but is radially unstable when $1< R_{c}\leq 1.198$. For $m=2$ and $m=3$ we find
radially unstable circular orbits with radius in the ranges $2\sqrt{2}/3\leq
R_{c}\leq 1.075$ and $2/\sqrt{5}\leq R_{c}\leq 1$, respectively. In contrast,
circular orbits for $m=4$ are always stable under small radial perturbations. We
conjecture that models with $m\geq 5$ are also radially stable. Figure
\ref{epiciclica} shows the behavior of $\kappa^{2}$ as a function of $R_{c}$ for
$m=1,2,3,4$. Figure \ref{vertical}, showing the behavior of $\nu^{2}$,
illustrates the stability under vertical perturbations. We find the following
ranges of vertical instability: $0\leq R_{c}\leq 1$ for $m=1$
($\kappa^{2}=-3\pi/2$); $0\leq R_{c}\leq 0.943$ for $m=2$; $0\leq R_{c}\leq
0.688$ for $m=3$; $0\leq R_{c}\leq 0.604$ for $m=4$. We see that the range of
vertical instability decreases with $m$.

\begin{figure}
\epsfig{width=2.75in,file=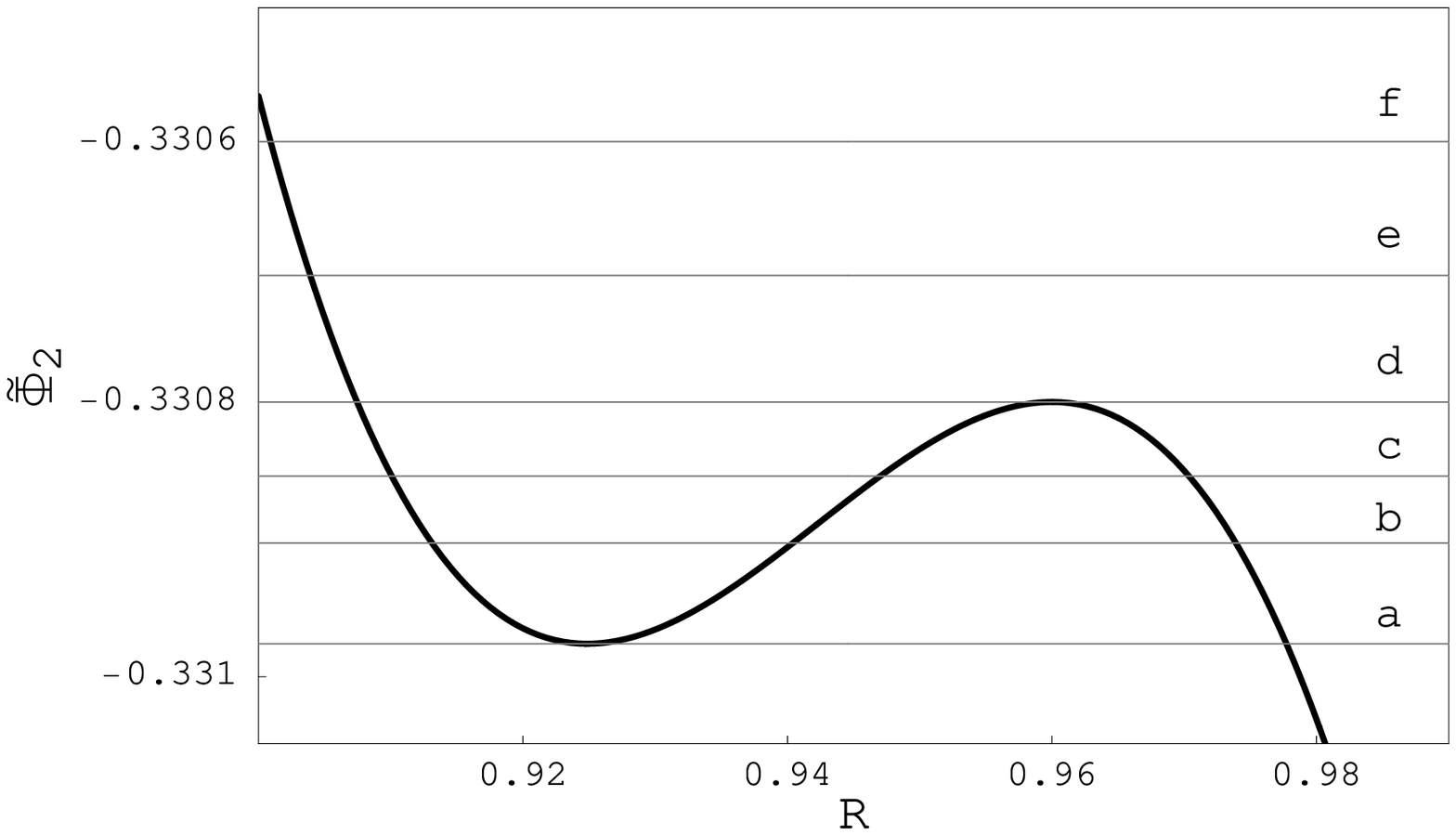}\\
  \caption{Effective potential for $m=2$ and $\ell=1.242$. Horizontal lines
corresponds to the following specific energy values: (a) -0.330975 (minimum of
$\tilde{\Phi}_{2}$), (b)-0.330900, (c) -0.330850, (d) -0.33079 (maximum of
$\tilde{\Phi}_{2}$), (e) -0.330700 and (f) -0.330600. Minimum and maximum of
$\tilde{\Phi}_{2}$ occur at $R_{c}=0.925$ and $R_{c}=0.960$.}\label{effective}
\epsfig{width=2.75in,file=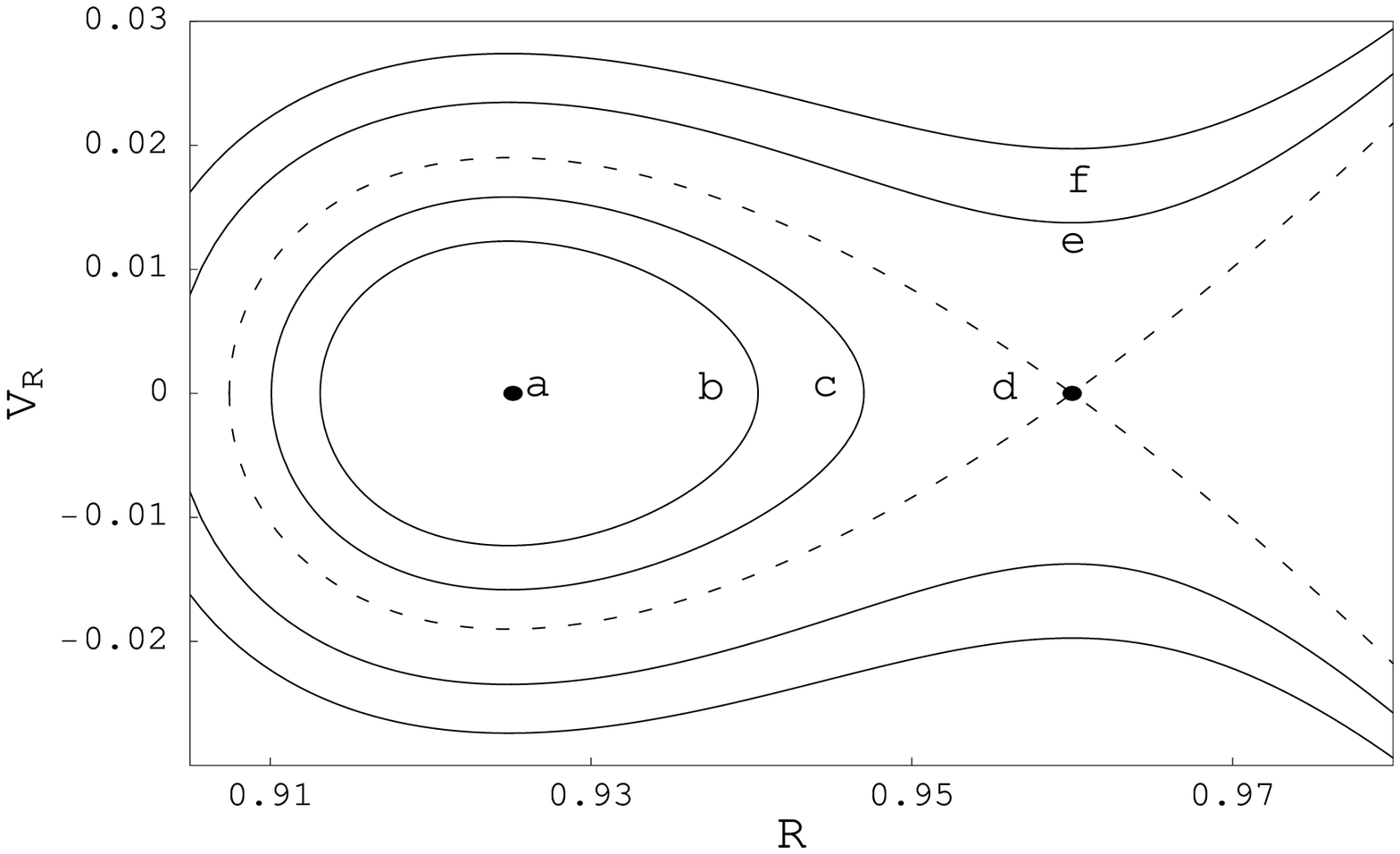}\\
  \caption{Phase portrait for $m=2$, corresponding to the same values of $\ell$
and $E$ showed in Figure \ref{effective}. Here (a) and (d) are stable and
unstable circular orbits, respectively. Curves (b) and (c) corresponds to
stable bounded motion, while (b) and (f) describe unbounded motion. The dashed
line represents a separatrix between bounded and unbounded motion
regions.}\label{phasecurve}
\end{figure}

General equatorial orbits are determined by (\ref{em}) together with the
conditions $\dot{V}_{z}=\dot{z}=z=0$ and $E=V_{R}^{2}/2+\tilde{\Phi}_{m}(R,0)$.
The motion is restricted by the inequality $E\geq \tilde{\Phi}_{m}(R,0)$ and, in
particular, we find bounded motion in a range $R_{1}\leq R\leq R_{2}$ if it
contains at least one critical value where $\tilde{\Phi}_{m}$ is minimum and
$\tilde{\Phi}_{m}(R_{1},0)\leq E\leq \tilde{\Phi}_{m}(R_{2},0)$. Figure
\ref{effective} shows the effective potential for $m=2$ and $\ell=1.242$, near
the disc edge. At energies (a), (b), (c), and (d) we have bounded orbits, and
(e), (f) corresponds to unbounded motion. In figure \ref{phasecurve} we present
the resulting phase portrait where the two regions of bounded and unbounded
motion are divided by a separatrix curve (dashed line). Similar phase portraits
can be performed for $m=1$ and $m=3$ if we set $\ell_{c}$ according to
(\ref{Lz}), for $1< R_{c}\leq 1.198$ and $2/\sqrt{5}\leq R_{c}\leq 1$,
respectively . For $m=4$ the effective potential does not present local maximums
and its phase portrait will not have any separatrix curve.

At a given energy $E$, determined by equation (\ref{Ecircular}) once we fix
$R_{c}$, the maximum possible value of the specific angular momentum is
$\ell_{c}$. Therefore, it is convenient to parameterize $\ell$ by means of the
ratio $k=\ell/\ell_{c}$ (see next section). It is useful to calculate the range
of values for $\ell_{c}$ such that $\tilde{\Phi}_{m}$ has a minimum in $0\leq
R\leq 1$, i.e. ensuring that bounded motion is always possible inside the disk.
In this way, we establish the limiting values of the integrals of motion for
which particles will never escape from the source. Using the relation
(\ref{Lz}), we found the following ranges for the specific axial angular
momentum: (a) $m=1$, $0 \leq |\ell_{c} |\leq \sqrt{3\pi}/2$; (b) $m=2$, $0 \leq
|\ell_{c} |\leq 2\sqrt{10\pi}/9$; (c) $m=3$, $0 \leq |\ell_{c} |\leq
\sqrt{21\pi/50}$; (d) $m=4$, $0 \leq |\ell_{c} |\leq 15\sqrt{7\pi}/64$.

\section{Disc-crossing orbits}\label{sec:cros}

In this section, we present numerical solutions of motion equations (\ref{em})
corresponding to bounded orbits outside the equatorial plane (except when they
cross the plane $z=0$). For certain values of $E$ and $\ell$, they are confined
to regions that contain the disc and will cross back and forth through it. As it
was showed by Hunter (\citeyear{Hunter}), this fact usually gives rise to many
chaotic orbits due to the discontinuity in the $z$-component of the
gravitational field, producing a fairly abrupt change in their curvatures. There
is an important exceptional case of this behavior: the Kuzmin's disc,
characterized by an integrable potential of the form
$\Phi=-GM[R^{2}+(a+|z|)^{2}]^{-1/2}$, with $a>0$. However, the so-called
Kuzmin-like potentials, characterized by $\Phi(\varepsilon)$ where
$\varepsilon=[R^{2}+(a+|z|)^{2}]^{1/2}$, are non-integrable and present the
behavior mentioned above. Generalized Kalnajs models present a very similar
structure and we can expect an analogous dynamics. Each potential
$\Phi_{m}(\xi,\eta)$ can be cast in a Kuzmin-like form if we take into account
that, according to (\ref{tco}), $\xi=(R_{+}+R_{-})/2a$ and
$\eta=(R_{+}-R_{-})/2ia$, where $R_{+}=[R^{2}+(z+ia)^{2}]^{1/2}$ and
$R_{-}=[R^{2}+(z-ia)^{2}]^{1/2}$. Moreover, they are characterized by a
z-derivative discontinuity in the disc, given by (Gonz\'alez \& Reina
\citeyear{GR})
\begin{equation}
\left(\frac{\partial\tilde{\Phi}_{m}}{\partial
z}\right)_{z=0^{+}}=-\left(\frac{\partial\tilde{\Phi}_{m}}{\partial
z}\right)_{z=0^{-}}=2\pi G\:\: \Sigma_{m} (R).\label{discont}
\end{equation}
Despite the above relation makes the KAM theorem inapplicable, we also found a
large variety of regular disc-crossing orbits.

\begin{figure}
\epsfig{width=2.75in,file=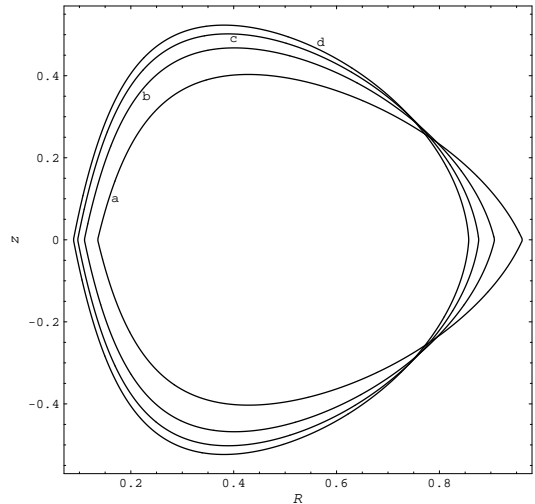}\\ \caption{Level contours of (a)
$\tilde{\Phi}_{1}$, (b) $\tilde{\Phi}_{2}$, (c) $\tilde{\Phi}_{3}$ and (d)
$\tilde{\Phi}_{4}$, when $E=-1.245$ and $\ell=0.2$.}\label{contornos}
\end{figure}

\begin{figure}
\epsfig{width=2.75in,file=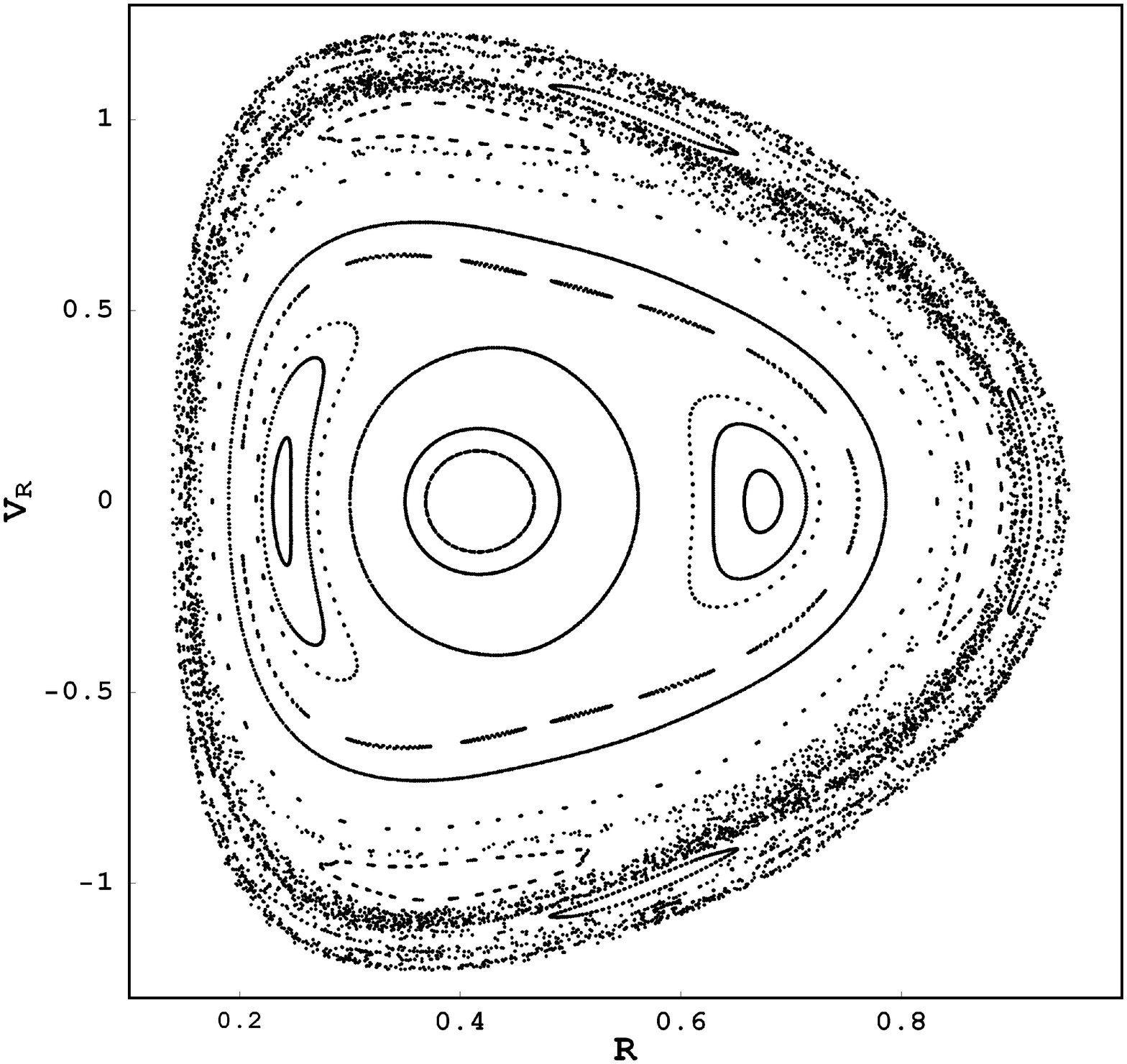}\\
\caption{Surface of section for $m=1$, $k=0.276$ and $E=-1.245$. We
have a
  small chaotic region with two resonant island chains and a central
  non-destroyed tori zone.}\label{kalnajs1}
\epsfig{width=2.75in,file=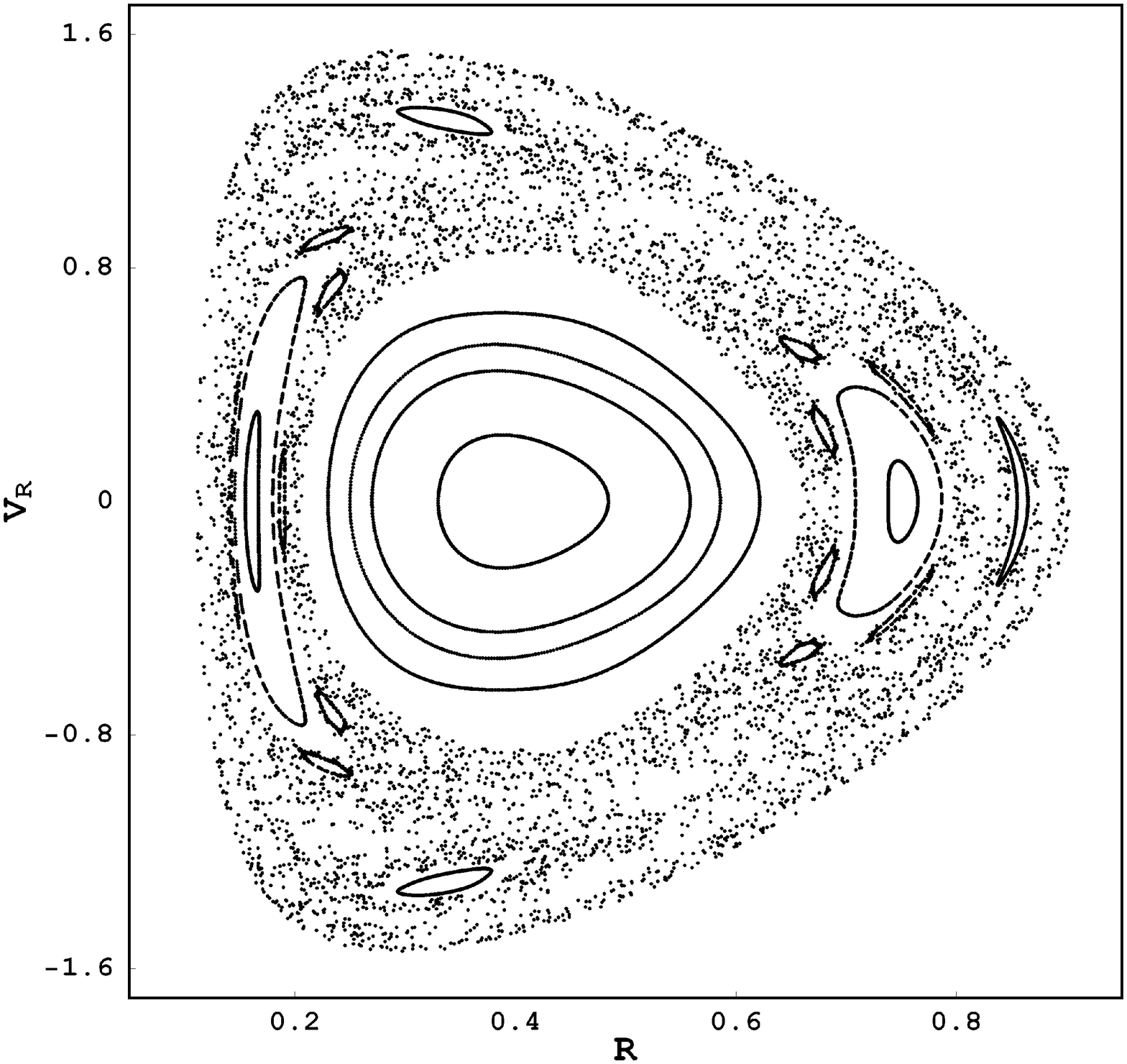}\\
  \caption{Surface of section for $m=2$, $k=0.266$, $E=-1.245$. The stochastic
  region is larger than in Fig. \ref{kalnajs1} and the non-destroyed tori zone
  is entirely formed by box orbits.}\label{kalnajs2}
\end{figure}

\begin{figure}
\epsfig{width=2.75in,file=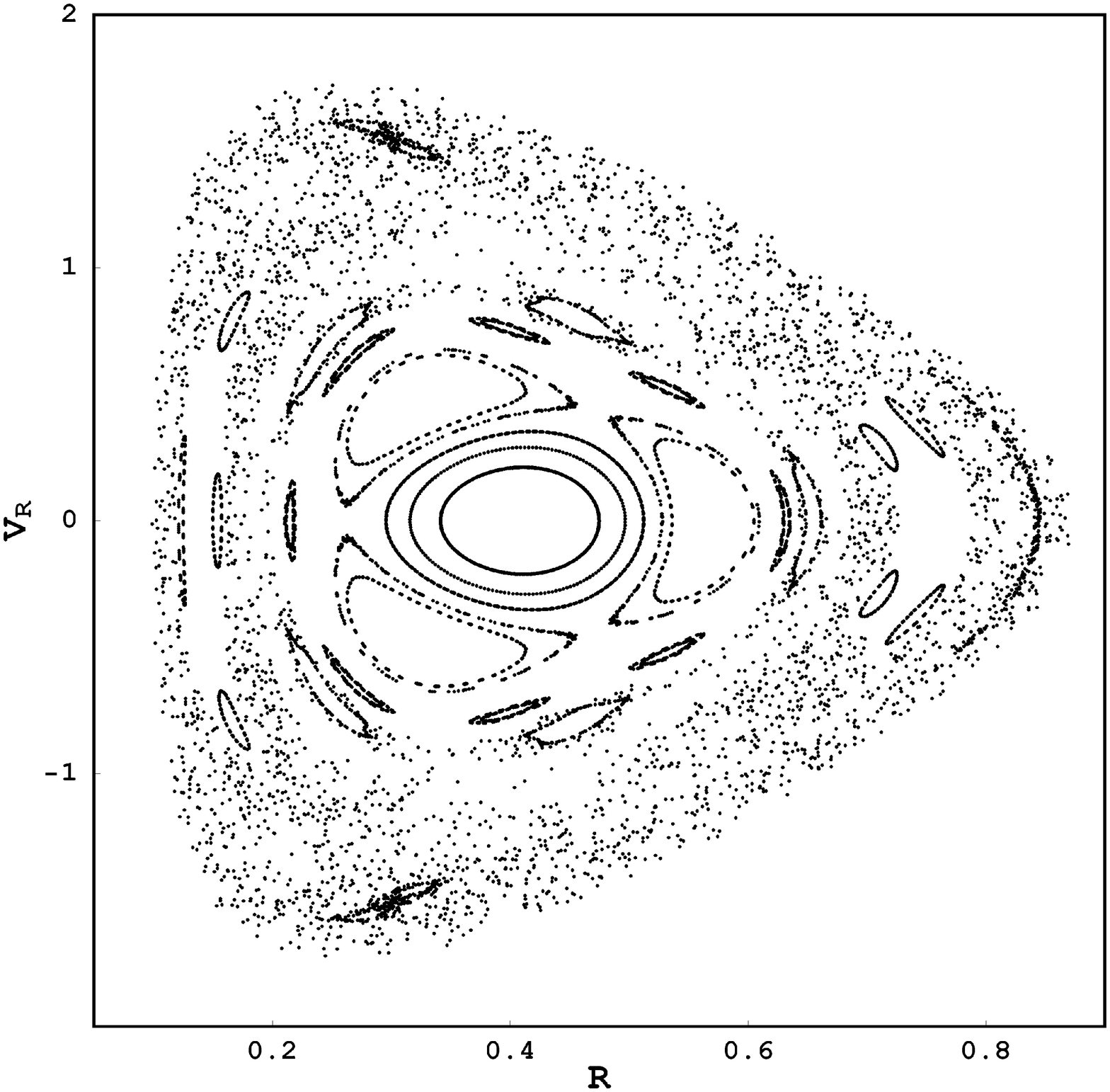}\\
  \caption{Surface of section for $m=3$, $k=0.263$ and $E=-1.245$.
  We see a prominent chaotic zone with
  island chains enclosing a regular region of box and loop orbits. The latter
  corresponds to the two central resonant islands.}\label{kalnajs3}
\epsfig{width=2.75in,file=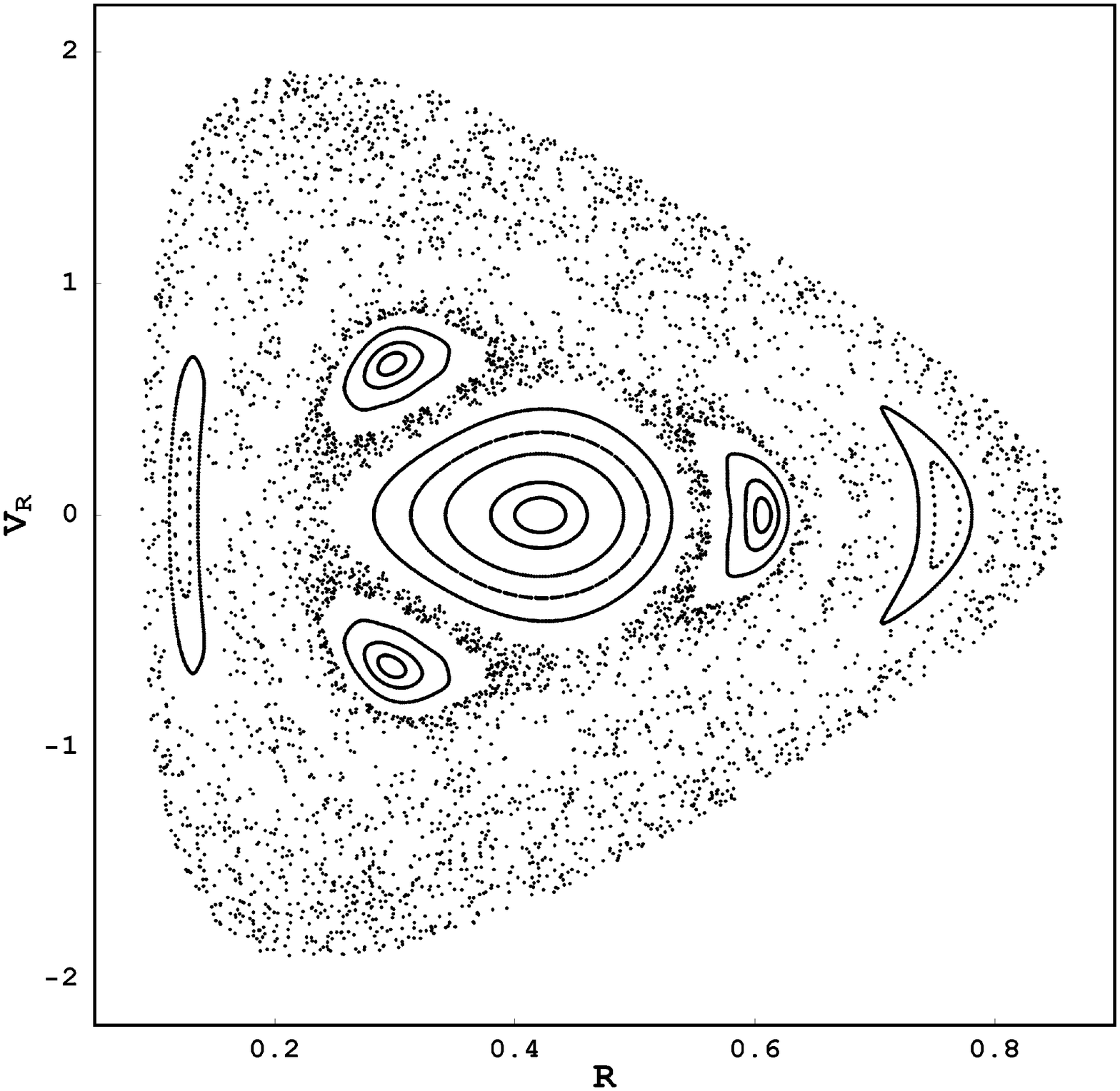}\\
  \caption{Surface of section for $m=4$, $k=0.262$ and $E=-1.245$. The chaotic
  region is larger than in the above figures and contain only two resonant
  island chains (the dotted curves at the extremum island are due to a periodic
  figure-of-eight orbit). The regular central zone is made entirely by box
  orbits.}\label{kalnajs4}
\end{figure}

In Figure \ref{contornos} we plot the level contours of $\tilde{\Phi}_{m}$ for
$m=1,2,3,4$, corresponding to $E=-1.245$ and $\ell=0.2$, i.e. $k=0.276, 0.266,
0.263$ and $0.262$, respectively. For this values, the motion of the particle is
confined to a region containing the disc. The corresponding $z=0$ surfaces of
section are shown in Figs. \ref{kalnajs1}-\ref{kalnajs4}, exhibiting a variety
of regular and chaotic trajectories. The Fig. \ref{kalnajs1}, that corresponds
to the values determining the contour (a) in Fig. \ref{contornos}, shows a large
KAM curve enclosing an island chain and three sets of three rings. There is also
a stochastic region with two island chains near section's edge. The large KAM
curve, the island chain and the central set of rings are produced by box orbits,
while the lateral rings as well as the last two island chains are formed by loop
orbits. The dotted curve, resulting from a banana boxlet periodic orbit, divide
the regular and stochastic region.

\begin{figure}
\epsfig{width=2.75in,file=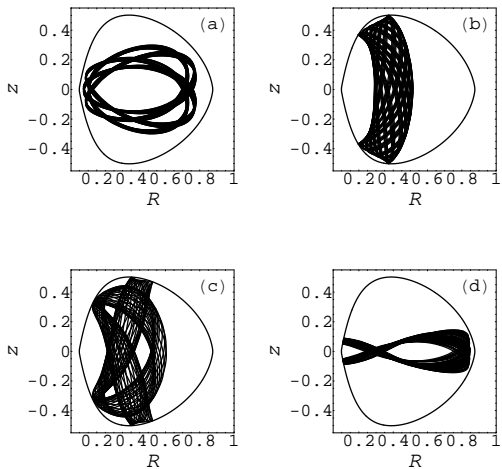}\\
  \caption{Orbits in the meridional plane for some initial conditions in Figure
  \ref{kalnajs3}: (a) Loop orbit at $R= 0.739172$, $V_{R} = 0.362539$ (middle
  island chain inside stochastic region); (b) Banana boxlet orbit at $R =
  0.421542$, $V_{R} = 0.362539$ (the largest central ring); (c) Pretzel shape
  loop orbit at $R= 0.405726$, $V_{R} = 0.573413$ (second island chain inside
  regular region); (d) A loop orbit at $R= 0.283155$, $V_{R} = 1.57197$ that
  forms the three denser zones near the section's edge. The outer boundary is
  the zero-velocity curve. } \label{orbitameridional3}
\end{figure}

Fig. \ref{kalnajs2} exhibits similar features as Fig. \ref{kalnajs1}. This
surface of section corresponds to the values defining the contour (b) shown in
Fig. \ref{contornos}. In this case we see a defined central region of box orbits
(four central rings) and an enclosing chaotic zone that contains a variety of
resonant islands of loop orbits. In this case the regions of box and loop orbits
are clearly separated, in contrast with Fig. \ref{kalnajs1} where they are
alternated. The surface of section corresponding to $m=3$ (contour (c) of Fig.
\ref{contornos}) is showed in Fig. \ref{kalnajs3}, exhibiting a regular region
composed by a central zone of banana boxlets and two resonant island chains of
loops. In the chaotic region we see island chains again and three denser zones
near the section's edge, formed by a loop orbit. Some of those meridional plane
orbits are plotted in Fig. \ref{orbitameridional3}. Finally, Fig. \ref{kalnajs4}
shows the surface of section for $m=4$, $k=0.262$ and $E=-1.245$. We find a very
prominent chaotic region with only two island chains and a small regular region
of box orbits.

\begin{figure}
\epsfig{width=2.75in,file=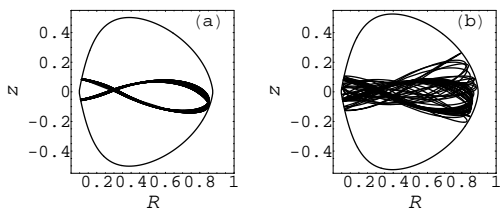}\\
  \caption{Orbits with the same $\ell$ and initial conditions as in Fig.
  \ref{orbitameridional3} (d). In the case (a), where $E=-1.225$, the orbit is
  still regular, whereas it turns chaotic when $E=-1.215$ in case (b). The
  outer boundary is the zero-velocity curve. } \label{orbit3-02evolution}
\end{figure}

The stochastic regions in Figs. \ref{kalnajs1}-\ref{kalnajs4} are due the
overlapping of many resonances caused by presence of the disc (Hunter
\citeyear{Hunter}). One can see this fact clearly in the three denser zones near
the section's edge of Fig. \ref{kalnajs3}, where the resonant islands are almost
overlapped at $E=-1.245$. When the energy increases to $-1.215$, for example,
the overlapping is complete and the trajectory turns to be irregular (Fig.
\ref{orbit3-02evolution}). In Fig. \ref{kalnajs4} we also note the footprint
traced by the overlapping of three prominent central islands.

\begin{table}
  \centering
  \begin{tabular}{|c|c|c|}
  \hline
  m & $LCN(\pm 0.0001)$ \\
  \hline
  1& $0.0108$ \\
  2& $0.0110$ \\
  3& $0.0118$ \\
  4& $0.0120$ \\
  \hline
\end{tabular}
  \caption{Estimation of the largest LCN for initial conditions $z=10^{-10}$,
  $R=0.681$,$V_{R}=0.819$ of a disc-crossing orbit at the chaotic region of
  Figs. \ref{kalnajs1}-\ref{kalnajs4}.}\label{Table}
\end{table}

In order to quantify the degree of instability of the orbits we calculate their
largest Lyapunov characteristic numbers (LCN), defined as
\begin{equation}
LCN=\lim_{\begin{array}{c}
          \Delta_{o}\rightarrow
0 \\
          t\rightarrow
\infty
        \end{array}
}\left[\frac{\ln(\Delta/\Delta_{o})}{t}\right],\label{lyapunov}
\end{equation}
where $\Delta_{o}$ and $\Delta$ are the deviations of two orbits at times $0$
and $t$ respectively. We obtain LCN using the procedure of Benettin et al
(\citeyear{Bennetin}). Thus by fixing the motion integrals as $E=-1.245$ and
$\ell=0.2$, choosing $\Delta_{o}\simeq 10^{-9}$ and $t=10^{7}$, we estimate $N$
corresponding to a typical chaotic disc-crossing orbit for the cases $m=1,2,3,4$
(Table \ref{Table}). We found that the degree of instability increases modestly
with $m$.

\begin{figure}
\epsfig{width=2.75in,file=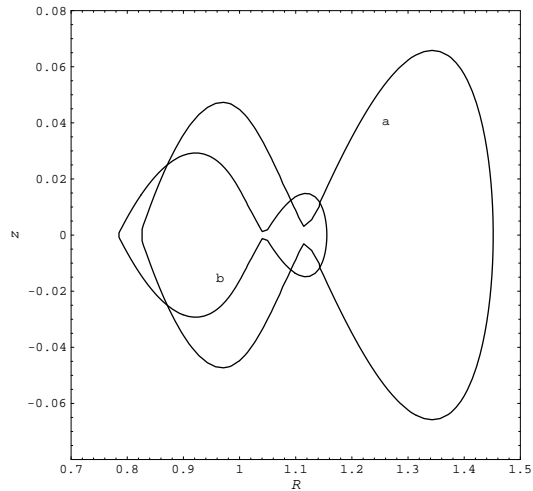}\\
  \caption{Level contours of (a) $\tilde{\Phi}_{1}$, $E=-0.335$, $\ell=1.287$;
(b) $\tilde{\Phi}_{2}$, $E=-0.389$, $\ell=1.196$. Both cases have two connected
regions, one of them containing the disc and the other is a disc-free zone.
}\label{contornos2}
\end{figure}
\begin{figure}
\epsfig{width=2.75in,file=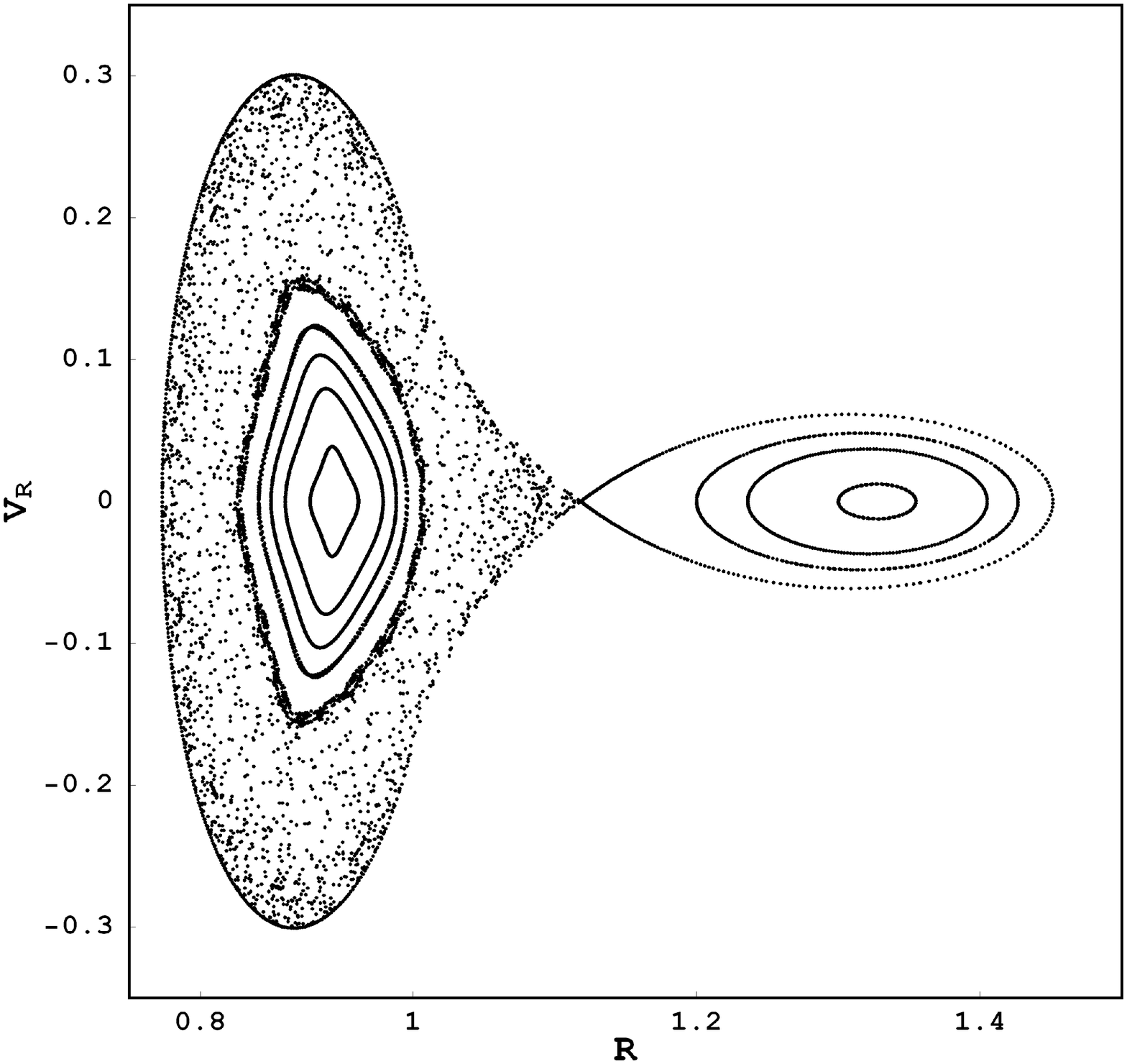}\\
  \caption{Surface of section for $m=1$, $\ell=1.287$ and $E=-0.335$ (contour
  (a) in Fig. \ref{contornos2}). There is a prominent chaotic zone of
  disc-crossing orbits to the left of $R=1.116$ (saddle point) and a regular
  region to the right.}\label{Akalnajs1}
\epsfig{width=2.75in,file=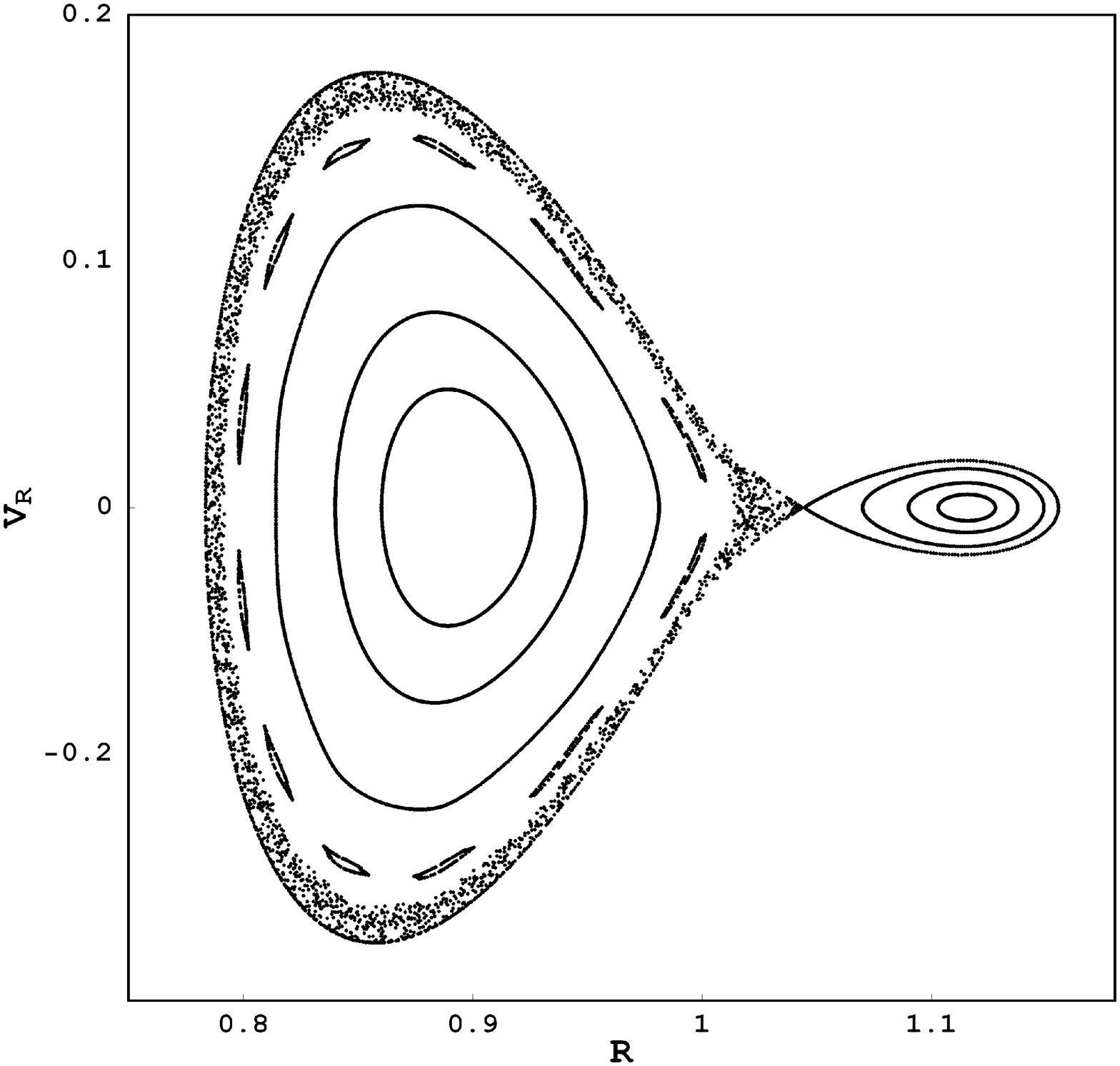}\\
  \caption{Surface of section for $m=2$, $\ell=1.196$ and $E=-0.389$ (contour
  (b) in Fig. \ref{contornos2}). The disc-crossing region has a narrow
  stochastic zone and a prominent KAM-curves region. We have again a small
  regular zone to the left of saddle point at $R=1.043$.}\label{Akalnajs2}
\end{figure}

An interesting phenomenon occurs when the effective potential has saddle points
outside the disc. This happens for the $m=1$ and $m=2$ discs, in whose case
there are critical points in the ranges  $1< R_{c}\leq 1.198$ and $1< R_{c}\leq
1.075$, respectively. Such equilibrium points are outside the disc but near its
edge, so for certain values of $E$ the contour of $\tilde{\Phi}_{m}$ will
contain only a fraction of the disc and a $z=0$ empty region. Then we will find
bounded disc-crossing and non-disc-crossing orbits. For example, we choose
$k=0.673$ (corresponding to $R_{c}=1.116$) and $E=-0.335$, obtaining the contour
(a) of Fig. \ref{contornos2}. The resulting surface of section (Fig.
\ref{Akalnajs1}) presents a large chaotic region to the left and a small totally
regular region to the right. Both are divided by the saddle point, so for
initial conditions near its left side the particle is ``trapped'' in the
stochastic zone. In contrast, for initial conditions near the right side of the
saddle point, the motion is confined to a region of non destroyed tori. In Fig.
\ref{Akalnajs2} we show a similar situation for $m=2$. This surface of section
corresponds to $k=0.991$ ($R_{c}=1.043$) and $E=-0.389$ (contour (b) of Fig.
\ref{contornos2}). We find a large zone of disc-crossing orbits with a small
chaotic component enclosing an island chain and regular rings of banana boxlets.
To the right of the saddle point there is a small regular region. For the cases
$m=3$ and $m=4$ it is not possible to obtain sections with the above features
since their corresponding effective potentials have not saddle points (the
product $\kappa^{2}\nu^{2}$ is positive in $R>1$).

\section{Concluding remarks}\label{sec.con}

One remarkable fact, suggested by the analysis \textbf{performed} in Sect. 2, is
that the stability of circular orbits under radial and vertical perturbations
increases with the parameter $m$ (for $m\geq 2$). More larger $m$ corresponds to
more stable models, when we deal with circular orbits. Another important fact is
that the range of axial angular momentum for which we can find bounded motion
inside the disc, decreases with $m$. Thus, loosely speaking, more and more
stable models have less and less possibilities to maintain their particles
inside the disc. These considerations have special relevance in the search of
equilibrium distribution functions (they are $E,\ell$-dependent) characterizing
such galaxy models.

The numerical calculations showed in Sect. 3 confirm the analysis made by Hunter
about disc-crossing orbits. There exist a chaotic motion induced by the presence
of the disk and, despite the current versions of KAM theorem do not apply (due
to the discontinuous force field), there is also a significant range of regular
orbits. Moreover, since in our case we deal with finite disc models, a
distinction between disc-crossing and non-disc-crossing orbits is sometimes
necessary. In the latter case we did not find chaotic motion, even in extreme
situations where there are saddle points outside the field source, and the disc
and disc-free region are connected (Figures \ref{Akalnajs1} and
\ref{Akalnajs2}).

Although one would be tented to think that such hyperbolic exterior points can
induce chaos at disc-free regions, what really happens is that the stochastic
motion tends toward disc regions and a completely regular motion is developed
outside there. These considerations have special relevance in galaxy models with
thin disc plus halo components. Particles belonging to the halo component will
follow a motion with the features mentioned above and, as it was showed by some
authors, this fact determines decisively the internal structure of such stellar
systems (see Ostriker, Spitzer and Chevalier (\citeyear{Ostriker})).

\section{Acknowledgments}

The authors thank to Leonardo Pach\'on for his valuable suggestions and
orientations. G. A. G. and F. L-S. want to thank the financial support from
COLCIENCIAS, Colombia, whereas that J. R-C. and F. L-S want to thank the
finantial support from {\it Vicerrector\'ia Acad\'emica}, Universidad Industrial
de Santander.


\begin{thebibliography}{9999}

\bibitem[\protect\citeauthoryear{Benettin}{1976}]{Bennetin} Benettin, G.,
Galgani, L. and Giorgilli A., 1972, Phys. Rev. A, 14, 2338.

\bibitem[\protect\citeauthoryear{Binney \& Tremaine}{1987}]{BT} Binney, J. and
Tremaine, S., 1987,  Galactic Dynamics. Princeton University Press, Princeton,
N. J.

\bibitem[\protect\citeauthoryear{Brandt}{1960}]{BR} Brandt, J. C., 1960, Ap. J.,
131, 211

\bibitem[\protect\citeauthoryear{Brandt \& Belton}{1962}]{BB} Brandt, J. C. and
Belton, M. J. S., 1962, Ap. J., 136, 352

\bibitem[\protect\citeauthoryear{Gonz\'alez \& Reina}{2006}]{GR} Gonz\'alez, G.
and Reina, J. 2006, MNRAS, 371 (4), 1873-1876.

\bibitem[\protect\citeauthoryear{Hunter}{1963}]{HUN1} Hunter, C., 1963, MNRAS,
126, 299

\bibitem[\protect\citeauthoryear{Hunter}{2005}]{Hunter} Hunter, C., 2005, Ann.
New York Acad. Sciences, 1045 (1), 120-138.

\bibitem[\protect\citeauthoryear{Kalnajs}{1972}]{KAL} Kalnajs, A. J., 1972, Ap.
J., 175, 63.

\bibitem[\protect\citeauthoryear{Kuzmin}{1956}]{KUZ} Kuzmin, G., 1956, Astron.
Zh., 33, 27

\bibitem[\protect\citeauthoryear{Martinet}{1971}]{Martinet1} Martinet, L. and
Hayli, A., 1971, A\&A, 14, 103.

\bibitem[\protect\citeauthoryear{Martinet}{1973}]{Martinet2} Mayer, F. and
Martinet, L., 1973, A\&A, 27, 199.

\bibitem[\protect\citeauthoryear{Martinet}{1974}]{Martinet3} Martinet, L., 1974,
A\&A, 32, 329.

\bibitem[\protect\citeauthoryear{Martinet}{1975}]{Martinet4} Martinet, L. and
Mayer, F., 1975, A\&A, 44, 45.

\bibitem[\protect\citeauthoryear{Ostriker}{1972}]{Ostriker} Ostriker, J. P.,
Spitzer, L., and Chevalier, R. A., 1972, ApJ, 176, L51.

\bibitem[\protect\citeauthoryear{Schmidt}{1956}]{Schmidt} Schmidt, M., 1975,
Bull. Astron. Inst. Neth, 13, 15.

\bibitem[\protect\citeauthoryear{Toomre}{1963}]{T1} Toomre, A., 1963, Ap. J.,
138, 385

\bibitem[\protect\citeauthoryear{Toomre}{1964}]{T2} Toomre, A., 1964, Ap. J.,
139, 1217

\bibitem[\protect\citeauthoryear{Wyse \& Mayall}{1942}]{WM} Wyse, A. B. and
Mayall, N. U., 1942, Ap. J., 95, 24

\end{thebibliography}
\end{document}